\documentclass[fleqn,usenatbib]{mnras}

\usepackage{newtxtext,newtxmath}
\usepackage[T1]{fontenc}
\usepackage{ae,aecompl}

\usepackage{xcolor}
\usepackage{graphicx}	% Including figure files
\usepackage{amsmath}	% Advanced maths commands
\usepackage{siunitx}
\usepackage[authoryear]{natbib}
\bibpunct{(}{)}{;}{a}{}{,}
\label{Bibliography}

\newcommand{\Msun}{\ensuremath{\,M_\odot}}
\newcommand{\Rsun}{\ensuremath{\,R_\odot}}

\newcommand{\yr}{\ensuremath{\,\mathrm{yr}}}

\newcommand{\kpc}{\ensuremath{\,\mathrm{kpc}}}

\newcommand{\startrack}{{\tt StarTrack}}

\renewcommand{\sci}[2]{\ensuremath{#1\times10^{#2}}}

\title[Predicting the self-lensing population in optical surveys]{Predicting the self-lensing population in optical surveys}

\author[Grzegorz Wiktorowicz]{
Grzegorz Wiktorowicz,$^{1,2,3}$\thanks{E-mail:gwiktoro@gmail.com}
Matthew Middleton$^{4}$, Norman Khan$^{4}$, \newauthor Adam Ingram$^{5}$, Poshak Gandhi$^{4}$ and Hugh Dickinson$^{6}$\\  
\\
    $^{1}$ National Astronomical Observatories, Chinese Academy of Sciences, Beijing 100101, China\\
    $^{2}$ School of Astronomy \& Space Science, University of the Chinese Academy of Sciences, Beijing 100012, China\\
    $^3$ Nicolaus Copernicus Astronomical Center, Polish Academy of Sciences, Bartycka 18, 00-716 Warsaw, Poland\\
    $^{4}$ Department of Physics \& Astronomy, University of Southampton, Southampton, SO17 1BJ, UK\\
    $^{5}$ Department of Physics, Astrophysics, University of Oxford, Denys Wilkinson Building, Keble Road, Oxford OX1 3RH, UK\\
    $^{6}$School of Physical Sciences, The Open University, Milton Keynes, MK7 6AA, UK\\
\\
}

\date{Accepted XXX. Received YYY; in original form ZZZ}

\pubyear{2020}

\begin{document}
\label{firstpage}
\pagerange{\pageref{firstpage}--\pageref{lastpage}}
\maketitle

% Abstract of the paper
\begin{abstract}
The vast majority of binaries containing a compact object and a regular star spend most of their time in a quiescent state where no strong interactions occur between components. Detection of these binaries is extremely challenging and only few candidates have been detected through optical spectroscopy. Self-lensing represents a new means of detecting compact objects in binaries, where gravitational lensing of the light from the visible component by the compact object produces periodic optical flares. Here we show that current and planned large-area optical surveys can detect a significant number ($\sim 100$--$10,000$s) of these self-lensing binaries and provide insights into the properties of the compact lenses. We show that many of the predicted population of observable self-lensing binaries will be observed with multiple self-lensing flares; this both improves the chances of detection and also immediately distinguishes them from chance-alignment micro-lensing events. Through self-lensing we can investigate long -- but previously hidden -- stages of binary evolution and consequently provide new constraints on evolutionary models which impact on the number and nature of double compact object mergers.
\end{abstract}


\section{Introduction}

The vast majority of binaries containing a neutron star -- NS or black hole -- BH, have been found through the action of mass transfer and accretion from a companion star, which typically requires the binary separation to be small \citep[see][]{Remillard0609}. Whilst X-ray binaries in the Galaxy number in the $\sim$thousands \citep[e.g.][]{vanHaaften1507}, this is orders of magnitude less than should be present in the form of non-interacting binaries with large orbital separations \citep{Wiktorowicz2006}. As non-accreting systems contain relatively pristine compact objects (having never accreted any material since their formation) they offer profound insights into the nature of the formation process and of the evolution of the binary itself.  

Detecting non-interacting and wide binaries can be achieved through three primary methods, firstly through detecting the astrometric shift due to the orbit of the companion star \citep[e.g.][]{Gould0606}, secondly through detecting radial velocity variations of the companion star \citep[e.g.][]{Casares1401}, and thirdly by detecting the self-lensing signature \citep[e.g.][]{Gould9506}. The former two have been discussed at length in relation to high precision surveys such as GAIA or LAMOST \citep[e.g.][and references therein]{Wiktorowicz2006}, whilst here we focus on the self-lensing signature of wide and non-accreting binaries in our Galaxy. 

Micro-lensing by isolated BHs is a well-established field and has been a vital tool for constraining limits on the mass of massive compact halo objects (MACHOs) \citep[e.g.][]{Wyrzykowski1110}. Self-lensing (SL) is a specific form of micro-lensing which occurs when a compact object transits across the disc of its optically bright companion star \citep{Gould9506}. The relative increase in brightness derived from magnifying the apparent size of the companion depends on the binary separation, the mass and radius of the compact object, and inclination to the observer \citep[e.g.][]{Witt9408}. In addition, the shape of the lensing flare and astrometric shift of the Einstein ring will depend on the spin (angular momentum) of the compact object \citep[e.g.][]{EbrahimnejadRahbari0508} which, in turn, encodes the physics of the formation mechanism. SL can also occur in super-massive BH binaries or in pre-merger double compact objects when one of the compact objects is accreting mass and produces X-ray radiation \citep{DOrazio1803,DOrazio2001,Ingram2021}, but in this study we focus only on optical observations.

Self-lensing has been observed in five binaries to-date, all containing white dwarfs \citep[WD][]{Kruse1404,Kawahara1803,Masuda1908}. In the case of WD lenses, the binary separation must be large enough such that the Einstein radius ($r_{\rm E}$) is larger than the WD's radius. Whilst this limits the separation and therefore orbital period ($P_{\rm orb}$) -- which also limits the Einstein crossing time ($\tau_{\rm E}$) and recurrence time ($t_{\rm rec}$) -- for WD systems, for those containing BHs and NSs, the binary separation can be considerably smaller (leading to a shorter period and crossing time). This opportunity for shorter periods yields a larger number of SL events for a given BH/NS system which, in turn, can distinguish the event from standard micro-lensing and provides opportunity for detailed follow-up (which is not as accessible for micro-lensing).

Whilst it is reasonable to expect many {\it potential} SL systems to be present in the Galaxy, their {\it detection} depends on the sensitivity, coverage and cadence (i.e. time between consecutive observations of the same field) of a given optical survey. In this paper we present a critical analysis of the SL population of both BH and NS binaries obtained via population synthesis under various initial conditions, using the grid of models prepared in \citet{Wiktorowicz1911}. We also predict the numbers and nature of SL systems we should expect to locate in the current and forthcoming data from optical surveys, focusing on the Transiting Exoplanet Survey Satellite
\citep[TESS;][]{Ricker1501}, the Zwicky Transient Facility \citep[ZTF;][]{Masci1901} and the Vera Rubin Observatory \citep[LSST;][]{Ivezic1903}.

\section{Methodology}\label{sec:methods}

\subsection{Self-lensing model}

\begin{figure}
	\includegraphics[width=\columnwidth]{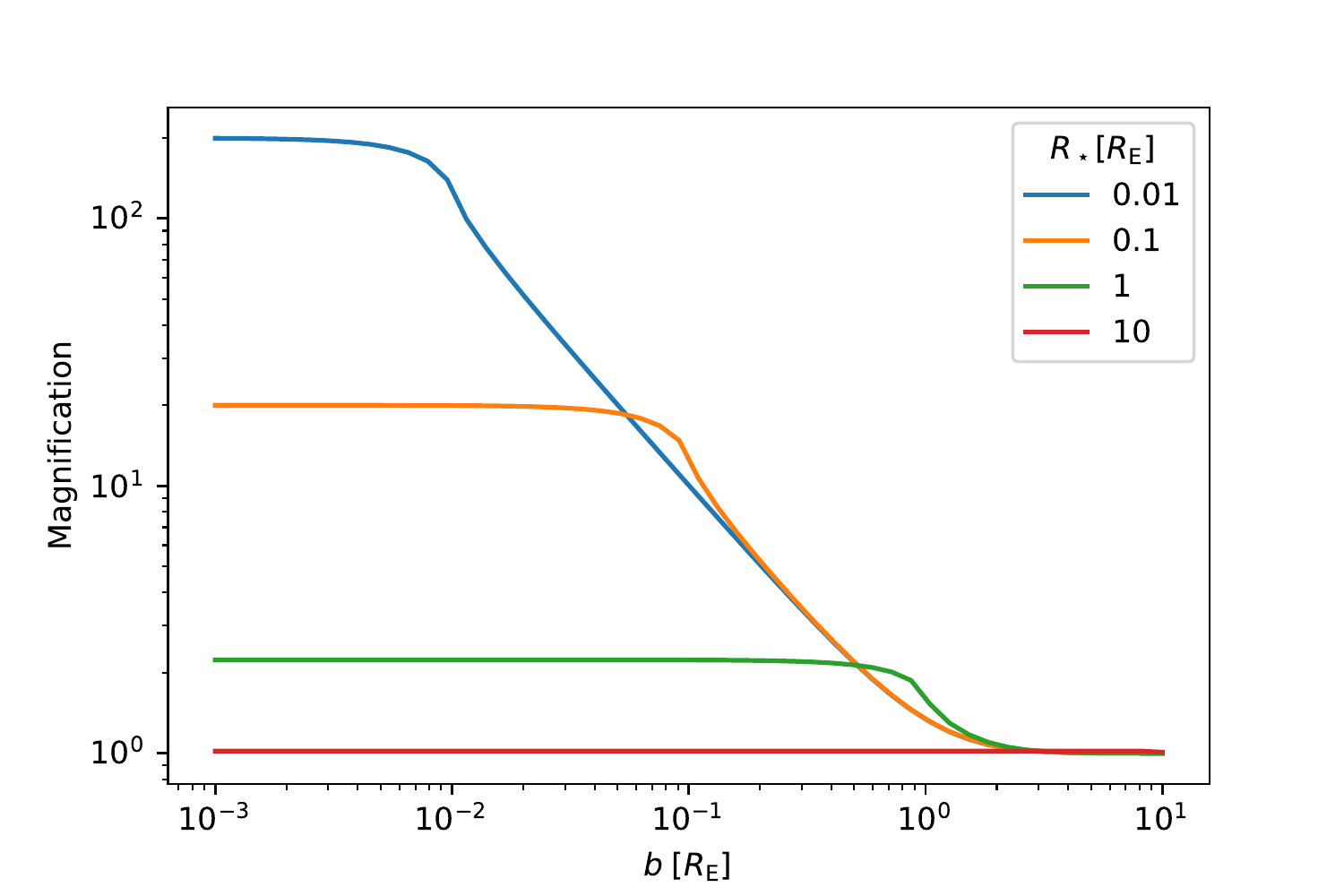}
    \caption{The relation between the \emph{peak} magnification and the impact parameter for a few relative sizes of the optical companion (source) star. The same relation holds between the \emph{instantaneous} magnification and the separation between the compact object and the centre of the source star in the lens plane.}
    \label{fig:magnification}
\end{figure}

The detailed SL signal for a given binary requires the application of ray-tracing codes within the appropriate metric. However, we can already obtain insights by assuming the simple analytical case where, in the absence of any limb-darkening and ignoring the eclipse of the companion star the maximal lensed brightness is magnified by a factor equal to \citep{Witt9408}
\begin{equation}\label{eq:mu_sl}
    \mu_{\rm SL} = \frac{1}{\pi} \left[ c_F F(k) + c_E E(k) + c_\Pi \Pi(n , k) \right],
\end{equation}
\noindent where $F$, $E$ and $\Pi$ are complete elliptic integrals of the first, second and third kind respectively. The parameters are:
\begin{eqnarray}
c_F &=& - \frac{b-r}{r^2} ~\frac{4 + (b^2-r^2)/2}{ \sqrt{ 4 + (b-r)^2 } } \nonumber \\
c_E &=& \frac{b+r}{2r^2} \sqrt{ 4 + (b-r)^2 } \nonumber \\
c_\Pi &=& \frac{2(b-r)^2}{r^2 (b+r)} \frac{ 1+r^2 }{ \sqrt{ 4 + (b-r)^2 } } \nonumber \\
n &=& \frac{4 b r}{(b+r)^2} \nonumber \\
k &=& \sqrt{ \frac{4n}{4+(b-r)^2} }.
\end{eqnarray}
\noindent where $b=(a/R_{\rm E})\cos i$ is the impact parameter (the minimum projected separation on the image plane between the centres of the two binary components) in units of $R_{\rm E}$, where $i$ is the inclination of the binary, and $r=R_\star/R_{\rm E}$ where $R_\star$ is the radius of the optical source (i.e. the companion star). The Einstein radius ($R_{\rm E}$) is defined as:
\begin{equation}
    R_{\rm E}=\sqrt{\frac{4GM_{\rm CO}}{c^2}\frac{D_{\rm LS}D_{\rm L}}{D_{\rm S}}},
\end{equation}
\noindent where $G$ is the gravitational constant, $M_{\rm CO}$ is the compact object (BH or NS) mass, and $D_{\rm LS}$, $D_{\rm L}$, and $D_{\rm S}$ are the distance between lens and source, the distance between observer and the lens, and the distance between the observer and the source, respectively. In the case of SL, $D_{\rm L}\approx D_{\rm S}$, therefore,
\begin{equation}
    R_{\rm E}\approx\sqrt{\frac{4GM_{\rm CO}a \sin i}{c^2}},
\end{equation}
\noindent where $a \sin i = D_{\rm LS}$ is the binary separation projected on the line of sight. The relationship between the maximum magnification and the impact parameter is presented in Figure~\ref{fig:magnification}, which can also be interpreted as the instantaneous magnification as a function of the separation between the Einstein radius and the centre of the source star in Einstein radii. We note that the formulae used in \citet{Agol0309} and \citet{Masuda1910} are simplifications of Equation~\ref{eq:mu_sl} for $R_{\rm E}\ll R_\star$.

All of the input parameters used to calculate $\mu_{\rm sl}$ are intrinsic to a given binary system except for the impact parameter $b$ which depends on the orientation of the orbit in space. We assume that all orbital orientations are equally probable and that a detectable SL event may occur only when $b<b_{\rm max} = 1+R_\star/R_{\rm E}$, i.e. the minimal separation between the compact object and the centre of the source star is smaller than $R_{\rm E}+R_\star$. However if the source is large ($R_{\star}>R_{\rm E}$), the magnification can be small ($\mu_{\rm sl}\lesssim2$; see Figure~\ref{fig:magnification}). The probability that a randomly orientated orbit will have such an impact parameter is $P_{\rm b} = P(b<1+R_{\rm E}/R_\star) = (R_{\rm E}+R_\star)/a$. In order to randomize the distribution of impact parameters, we calculate them as $b=a\cos i/(R_{\rm E}+R_\star)$, where $\cos i$ comes from a uniform distribution between $0$ and $(R_{\rm E}+R_\star)/a$.

In micro-lensing, the Einstein crossing time is defined to be the maximum time (i.e. for an edge-on orientation) for which the separation on the sky between the source and lens is smaller than $R_{\rm E}$. In the case of SL (where $D_{\rm L}\approx D_{\rm S}$), the optical star can no longer be interpreted as point-like, therefore, we extend the definition of the crossing time to include objects where at least a fraction of the companion star's disk, projected on the lens plane, falls within the Einstein radius from the lens, which we call an effective crossing time ($\tau_{\rm eff}$). Additionally, by allowing for non-zero impact parameters, we obtain shorter, but more realistic, crossing times than found in previous studies \citep[e.g.][]{Masuda1908}. Specifically, we calculate the effective crossing time as: 
\begin{equation}\label{eq:tau_eff}
    \tau_{\rm eff}=\frac{P_{\rm orb}(R_{\rm E}+R_\star)}{\pi a\sin i}\sqrt{1-\left(\frac{b}{b_{\rm max}}\right)^2},
\end{equation}
\noindent where $P_{\rm orb}$ is the orbital period. We used a mean value for the distance between the stars ($<r> = a$) and a mean value for the orbital velocity ($<v> = 2\pi a / P_{\rm orb}$). We note that $\tau_{\rm eff}$ is different to $\tau_{\rm sl}$ as defined in \citet{Masuda1910}, where the average value of the impact parameter was used.

\begin{table}
    \centering
    \begin{tabular}{lcccccc}
        \hline
        & $t_{\rm survey}$ [yr] & cadence [days] & $m_{\rm lim}$ & $m_{\rm sat}$ & $n_{\rm filters}$ & $m_{\rm lim,eff}$\\
        \hline
        ZTF & 5 & 1 & 21 & 12 & 2 & 21.93 \\
        LSST & 10 & 4.6 & 24 & 14 & 1 & 24.55\\
        TESS & 0.07$^a$ & 0.0014 & 12 & 4 & 1 & 12.55\\
        \hline
    \end{tabular}
    \caption{Key parameters for the survey instruments used in our calculations. $t_{\rm survey}$: duration of survey; $m_{\rm lim}$: magnitude limit for SNR = 5 and one filter; $m_{\rm sat}$: instrument saturation limit; $n_{\rm filters}$: number of observing filters used; $m_{\rm lim,eff}$: corrected magnitude limit. $^a$ the TESS mission duration is actually $2$ years but every field will be observed for $27$ days. We note that a fraction of the stars will be observed for longer than this due to field intersection, but we do not consider this in our results (therefore our results are somewhat conservative).} 
    \label{tab:instruments}
\end{table}

\begin{figure*}
    \centering
    \includegraphics[width=\textwidth]{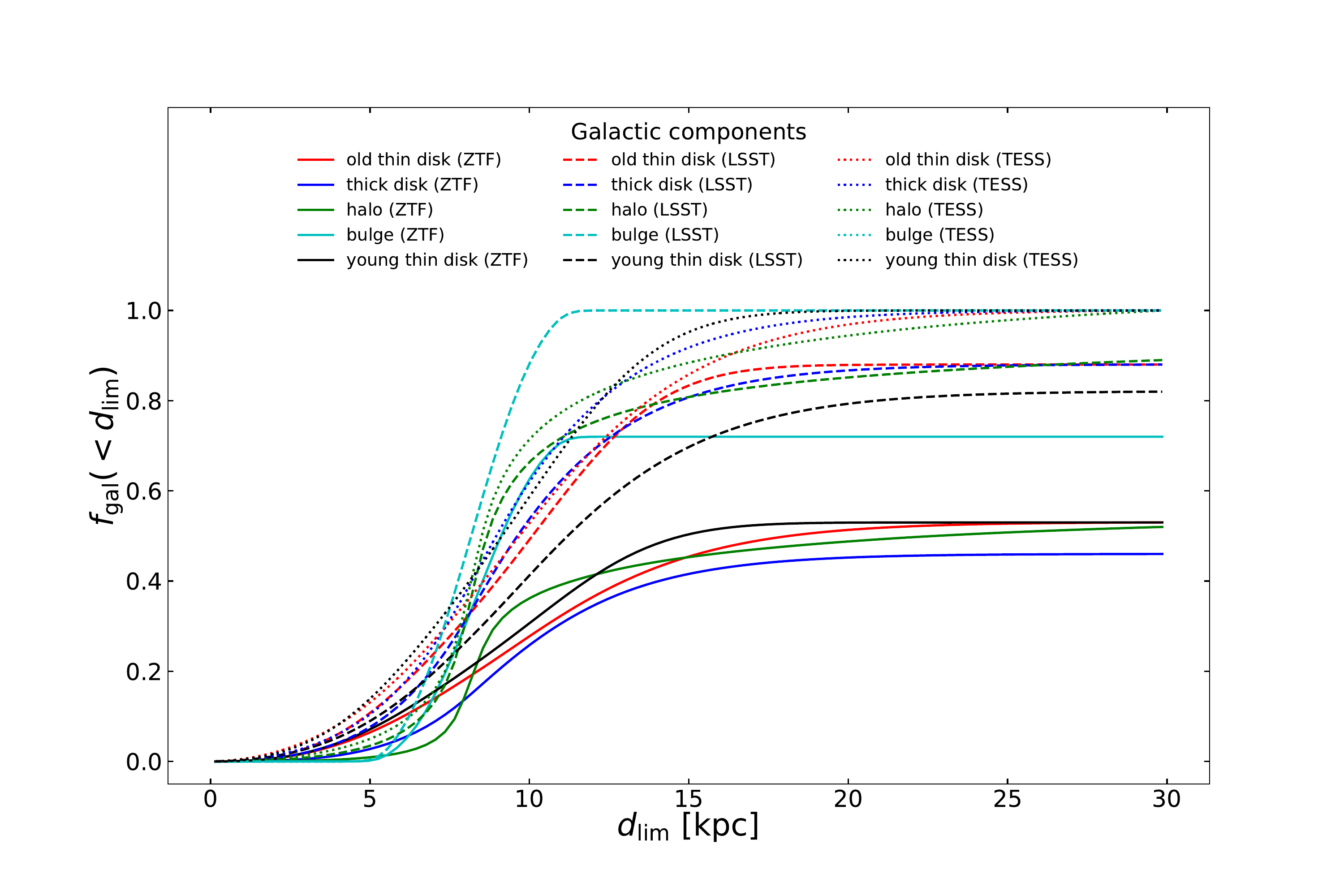}
    \caption{Density profiles showing the fraction of the Galactic component's mass in relation to distance from the Earth within the field of view of different optical surveys. The spatial distribution of stellar mass in the Galaxy follows \citet{Robin0310} as described in \citet{Wiktorowicz2006}.}
    \label{fig:profiles}
\end{figure*}

\subsection{Binary population synthesis}

This work uses the publicly available database of BHs in synthetic stellar populations \citep{Wiktorowicz1911}\footnote{\url{https://universeathome.pl/universe/bhdb.php}}. This catalogue was obtained through use of the {\tt startrack} population synthesis code \citep{Belczynski0801,Belczynski2004} and provides ready-to-use data for a variety of studies. We have extended the database to include NS progenitors by including binaries with primary masses on the ZAMS as low as $5\Msun$. We have also utilised the reference model for binary population synthesis as defined in \citet{Wiktorowicz1911} together with the initial mass function (IMF) of $P(M_{\rm ZAMS})\propto M_{\rm ZAMS}^\Gamma$, where $\Gamma=-2.3$ for the standard (std) model, $\Gamma=-1.9$ for the flatIMF model, and $\Gamma=-2.7$ for the steepIMF  model. For these models, we utilise variations in metallicity (models with low-Z: $1\%$ solar, and mid-Z: $10\%$ solar) to account for the chemical diversity in the Milky Way. We refer the reader to \citet{Wiktorowicz1911} for additional details and a thorough discussion of the models and simulations.

For our analysis, we have chosen those binaries containing compact objects (NS/BHs only) and which are evolving through a non-interaction phase and explicitly ignore those which are undergoing mass transfer (noting that these systems may also self-lens). From the population synthesis calculation, we have all of the requisite ingredients to simulate the SL signatures for binaries in our sample (using the SL model described above) and determine the proportion which might be discovered by various ongoing and forthcoming surveys. We make the simplifying assumption that the peak brightness of the SL event is maintained for the entire Einstein crossing time (i.e. that the ingress and egress of the lensing is much smaller than $\tau_{E}$). This assumption only breaks down for very small stars with radius comparable to that of the lens.

\subsection{Observational predictions}

To account for the stellar mass distribution in the Galaxy, we use the Milky Way model of \citet{Robin0310}, modified as described in \citet{Wiktorowicz2006}. Following the procedure introduced in \citet{Wiktorowicz2006}, which improves on the traditional Monte Carlo-based sampling methods, we calculated the fraction of the Galactic mass within a distance $d$, observable from the Earth for the instruments under consideration (Figure~\ref{fig:profiles}). To calculate bolometrically corrected absolute magnitude ($m_{\rm abs}$) we use the YBC database \citep[see also Sec~\ref{sec:instruments}]{Chen1912}. We employed the \citet{Spitzer78} scaling of interstellar extinction, $A=1 \mathrm{mag}/\mathrm{kpc}$. Consequently, the apparent magnitude of a flare\footnote{In this work, by the luminosity of the flare we understand the difference between the luminosity of the magnified ($\mu>1$) and not magnified ($\mu = 1$) source.} is given by:
\begin{equation}\label{eq:m_app}
    m_{\rm app, flare}=m_{\rm abs}-2.5\log_{10}\left[(\mu_{\rm sl}-1)/\left(\frac{d}{\rm kpc}\right)^{2}\right]+\left(\frac{d}{\rm kpc}\right)+10.
\end{equation}

In our sample we include only systems in which a single SL flare gives a signal-to-noise ratio $\mathrm{SNR} > 3$. In the case of ZTF which provides observations in two filters, we allowed for stacking of observations which improves the SNR. The effective magnitude limit is calculated as:
\begin{equation}\label{eq:m_corr}
    m_{\rm lim,eff}=m_{\rm lim}-2.5\log_{10} \frac{\rm SNR}{\rm SNR_{\rm ref}} n_{\rm filter}^{-\frac{1}{2}},
\end{equation}
\noindent where $m_{\rm lim}$ is the magnitude limit for the given instrument if  $\mathrm{SNR}=\mathrm{SNR}_{\rm ref}$ ($=5$ in our case, i.e. $5\sigma$ significance level) and $n_{\rm filter}$, which is the number of filters used, is equal one. Table \ref{tab:instruments} presents the parameters used in this study for our chosen instruments. We note that the choice of a smaller SNR is motivated also by the fact that in SL we can co-add separate flares to increase data coverage of a flare profile, so the SNR will improve with time.

For each system in our simulated sample, we use Eq.~\ref{eq:m_corr} to calculate the distance out to which its flare is observable ($d_{\rm max}$) from the implicit function $m_{\rm app, flare}(d=d_{\rm max})=m_{\rm lim}$. Similarly, we include the effects of saturation ($m_{\rm sat}$ in Table~\ref{tab:instruments}) by calculating the distance at which the lensed star is not saturated ($d_{\rm sat}$), from the implicit relation $m_{\rm app, flare}(d=d_{\rm sat})=m_{\rm sat}$. Additionally, we place a limit on the smallest observable distance by ensuring that $d_{\rm min}$ is not smaller than $0.1\kpc$ through $d_{\rm min}=\min(0.1\kpc, d_{\rm sat})$ which avoids including any objects closer than the closest stars ($0.1\kpc$ roughly corresponds to the distance to the nearest super-giants, $\alpha$ UMi Aa: $\sim0.1$ kpc \citep{Turner1301}, Antares: $0.17\kpc$ \citep{vanLeeuwen0711}, and Betelgeuse: $0.22\kpc$ \citep{Harper1707}). As it turns out, even for a continuous mass-distribution model, the fraction of stars inside $d_{\rm min}$ is negligible for a typical case and doesn't influence our results. Following this calculation, the distributions in Fig.~\ref{fig:profiles} are used to obtain the fraction of the stellar mass in the Galaxy, found within the distance at which the flare is observed by the instrument of choice. As such, the probability that a system will be present in that region of space is proportional to the stellar density in that Galactic location and is given by
$f_{\rm d}=f_{\rm gal}(<d_{\rm max})-f_{\rm gal}(<d_{\rm min})$.
Finally, the probability that a random source in the simulation results will undergo an observable SL event at the moment of observation can be calculated to be:
\begin{equation}
    P_{\rm obs}=f_{\rm d}f_{\rm age}f_{\rm SF}P_{\rm b},
\end{equation}
\noindent where $f_{\rm age}=\mathrm{SFR}(t=t_{\rm age})dt$ is the probability that the system is observed currently, SFR$(t)$ is the star formation history (i.e. the star formation rate as a function of time), $t$ is the look-back time, $t_{\rm age}$ is the age of the system (time since ZAMS), and $f_{\rm SF}$ is a scaling factor between the simulated stellar mass and the Milky Way's stellar mass. $f_{\rm SF}$ depends on the IMF and in our case equals $168$/$295$/$408$ for steep/standard/flat ($\Gamma=-2.7/-2.3/-1.9$) IMFs, respectively. We note that the typical properties of binaries harbouring COs may differ significantly from the general stellar population in the Galaxy due to the natal kicks that COs may obtain during birth. Natal kicks not only influence the spatial distribution of CO binaries, but also binary parameter distributions \citep[e.g. the $P_{\rm orb}$; see][]{Gandhi2005}. Nonetheless, these distributions are poorly sampled and in this study we treat the distributions for the general stellar population as adequate.

The ability to observe a given SL flare is further influenced by its duration (of the order of the Einstein crossing time) as given by equation \ref{eq:tau_eff}, $P_{\rm orb}$, and the parameters of the survey considered. Each survey has a duration $t_{\rm survey}$ -- which is either the total duration (ZTF, LSST) or the time spent observing a particular field (TESS) -- and a cadence of $t_{\rm c}$, such that a given field is observed $N=\lfloor t_{\rm survey}/t_{\rm c} \rfloor$ times altogether. Since the duration of each observation is far shorter than the typical flare duration, we assume that each survey consists of $N$ instantaneous snapshots spaced by $t_{\rm c}$. For SL flares with duration longer than the survey cadence, $\tau_{\rm eff}>t_{\rm c}$, $N_{\rm rec}= t_{\rm survey}/P_{\rm orb}$ flares will be observed by the survey and each flare will be covered by an average of $N_{\rm points} = \tau_{\rm eff}/t_{\rm c}$ observations. In this case, the probability that at least one flare is seen from a given self-lensing source by a given survey is therefore $P_{\rm vis} = {\rm min}[1,t_{\rm survey}/P_{\rm orb}]$. We see that a survey with $t_{\rm survey} > P_{\rm orb}$ is sure to see at least one flare in this $\tau_{\rm eff}>t_{\rm c}$ limit, whereas $P_{\rm vis}<1$ if the entire survey only covers a fraction of an orbital period.

In the opposite limit of $\tau_{\rm eff} < t_{\rm c}$, each SL flare that occurs during the survey will either be missed completely or will be covered by only one observation ($N_{\rm points}=1$). The probability of a single observation occurring during a lensing flare is $\tau_{\rm eff}/P_{\rm orb}<1$. We employ the following `brute force' method to determine the effective visibility of these systems. We assume without loss of generality that the flare occurs at an orbital time of $[P_{\rm orb}-\tau_{\rm eff},P_{\rm orb}]$; we then choose an initial (i.e. at the beginning of the survey) time during the system's orbit, $t_0$, from a uniform distribution between $0$ and $P_{\rm orb}$. If $t_0\in[P_{\rm orb}-\tau_{\rm eff},P_{\rm orb}]$ then we treat the flare as being detected. Otherwise, we increment the time $t$ (equal initially to $t_0$) by the observing cadence until $t\geq t_{\rm survey}$. If $\exists_{n<\lfloor(t_{\rm survey}-t_0)/t_{\rm c}\rfloor} \left[ t_0 + n t_{\rm c} \right]_{\mod(P_{\rm orb})} \in[P_{\rm orb}-\tau_{\rm eff},P_{\rm orb}]$, then we treat the flare as detected ($P_{\rm vis}=1$). Simultaneously, we count the number of visible flares (i.e. the number of different values of $n$ for which $\left[ t + n t_{\rm c} \right]_{\mod(P_{\rm orb})} \in[P_{\rm orb}-\tau_{\rm eff},P_{\rm orb}]$) as $N_{\rm rec}$, and calculate $N_{\rm points}$ as $\tau_{\rm eff}/t_{\rm c}$. Note that for $t_{\rm survey} \gg P_{\rm orb}$, we can simply write $N_{\rm rec}\approx(\tau_{\rm eff}/P_{\rm orb}) (t_{\rm survey}/t_{\rm c})$, but out of this limit, resonance between $t_{\rm c}$ and $P_{\rm orb}$ can make the actual $N_{\rm rec}$ significantly lower than the naively expected value; our brute force method accounts for such resonances.

The above calculations are repeated for all evolutionary time-steps (typically, $10$--$1000$ per system) and for all presently non-interacting systems harbouring a BH or a NS and a visible companion within the database. The expected number of observed SL events per evolutionary time-step equals: 
\begin{equation}
    E(\xi_{\rm SL})=P_{\rm obs}P_{\rm vis},
\end{equation}
\noindent where $P_{\rm obs}$ and $P_{\rm vis}$ are calculated for instantaneous system parameters during the time step. The total expected number of SL events is calculated as:
\begin{equation}
    E(n_{\rm SL})=\sum E(\xi_{\rm SL}),
\end{equation}
\noindent where the summation is over all systems and all time steps. We now consider the numbers of detectable SL systems accessible to some of the major ongoing and forthcoming optical surveys.

\subsection{Optical surveys}\label{sec:instruments}

We have chosen three optical surveys with which to calculate predictions based on the models and approach described above. The survey parameters needed for the analysis are collected in Table~\ref{tab:instruments}. In our case, the bolometric corrections are added early during the calculations, before the apparent magnitude limit or the saturation are included. Below, we briefly discuss the surveys.

\subsubsection{TESS}

The main goal of the Transiting Exoplanet Survey Satellite (TESS) is the discovery of extra-solar planets through detecting transits \citep{Ricker1408}, and will observe nearly the entire sky. TESS has a very high cadence ($2$ min\footnote{actually, the cadence is even higher but the individual observations are stacked into $2$ min frames}), which enhances the detection probability of close binaries and increases the number of detected SL flares as well as their respective data coverage. More than $\mathbf{2\times10^5}$ stars will be observed with such a cadence \citep{Sullivan1508}, whilst the remainder will be observed with a $30$ min cadence. Although the mission duration is $2$ years, each of the $26$ TESS fields will be observed for a consecutive $27$ days and, in some overlapping sections, a continuous coverage of up to a $1$ year for a fraction of the sky will be possible. Whilst the cadence is the highest of all of the surveys we consider here (which enables a large number of flares to be detected and thereby improves the chances of distinguishing the SL event from micro-lensing), the apparent magnitude limit of $m_{\rm lim}=12$ for $\mathrm{SNR}=5$ ($5\sigma$), or $12.55$ for $3\sigma$ is relatively restrictive. When calculating the numbers of SL systems detected by TESS, we assume an optimistic scenario that SL binaries will be mostly detected among the pre-selected group of target stars with $2$ min cadence, whilst we assume a total observation duration of $27$ days ($\approx0.07$ year) for all stars regardless of their location on the sky. TESS uses a red-optical bandpass covering the wavelength range of $\sim600$--$1000$ nm, whereas its saturation limit was assumed as $4$ mag.

\subsubsection{ZTF}

The Zwicky Transient Facility (ZTF) is presently performing two public surveys, the polar plane survey (PPS) and Galactic plane survey (GPS). The PPS is covering Galactic regions spanning declination $d>-31^\circ$ and galactic latitudes $|b|>7^\circ$, with observations in the g and r bands, and a $3$ day cadence. The GPS covers $d>-31^\circ$ similarly to PPS, but $|b|<7^\circ$, with observations in the g and r bands with a $1$ day cadence. In our calculations we assume that we can combine observations from the two filters (i.e. assuming the crossing time is long compared to the repeat observations) and effectively double the exposure time and thereby decrease the noise level. The single exposure $5\sigma$ apparent magnitude limit for ZTF is $21$ mag, but using two filters and lowering the detection threshold to $3\sigma$, we can increase it to $21.93$ (Eq.~\ref{eq:m_corr}, see Tab.~\ref{tab:instruments}). In this study, we have used only the GPS survey whose field-of-view includes a significantly larger fraction of the stellar mass in the galaxy ($\sim53\%$) than the PPS ($\sim2\%$). The saturation limit for ZTF is adopted as $12$ mag and the filter used for bolometric correction is {\it r}.

\subsubsection{Vera Rubin LSST}

The Vera C. Rubin Observatory \citep{Ivezic1903}, also known as the Large Synoptic Survey Telescope (LSST), is an optical survey telescope, which will observe the southern hemisphere (declination $\delta<0$) with a $4.6$ day cadence, down to a depth of $24$ or $24.55$ mag ($5\sigma$ or $3\sigma$, respectively). We note that this cadence is still not final for LSST \citep[see e.g.][]{Johnson1903} and may change in the future. The shorter the observing cadence, the higher the chance of detecting short SL events, and the better the data coverage of flares becomes; however technical limitations and observational strategies put limits on the cadence. The saturation limit for LSST is $14$ mag in our simulations and we assume observations in the {\it r} band for the bolometric correction. The LSST's first light is expected in 2023. Among other scientific goals, LSST will map the Milky Way down to its magnitude limit; as the majority of Galactic stars reside in the southern hemisphere, LSST's field-of-view (FOV) contains $90\%$ of the stellar mass in the Galaxy making it an ideal instrument for detecting SL events.

\section{Results}

\begin{table*}
    \centering
    \begin{tabular}{llcccccl}
        \hline
        Instrument & IMF & $n_{\rm sl}$ & $f_{1,1}$ & $f_{\rm NS}$ & $m_{\rm source,max}$ & $\mu_{\rm sl,max}$\\
        \hline 
       ZTF &        std & $4027.97$ & $0.45$ & $0.93$ & $24.40$ & $3554.11$\\
       ZTF &    flatIMF & $5529.63$ & $0.45$  & $0.88$ & $24.40$ & $3664.52$\\
       ZTF &   steepIMF & $2142.49$ & $0.46$ & $0.96$ & $24.50$ & $6140.31$\\
      LSST &        std & $10284.91$ & $0.78$ & $0.91$ & $24.20$ & $4401.18$\\
      LSST &    flatIMF & $13939.22$ & $0.79$ & $0.85$ & $24.60$ & $6585.01$\\
      LSST &   steepIMF & $5284.28$ & $0.78$ & $0.94$ & $24.60$ & $3509.54$\\
      TESS &        std & $123.28$ & $0.71$ & $0.95$ & $24.50$ & $10896.10$\\
      TESS &    flatIMF & $172.78$ & $0.69$ & $0.90$ & $24.40$ & $5607.77$\\
      TESS &   steepIMF & $64.59$ & $0.71$ & $0.97$ & $24.40$ & $4349.98$\\
        \hline
    \end{tabular}
            \caption{Predictions for self-lensing,  non-interacting (at present) binaries for ZTF, LSST, and TESS. Results were calculated for a realistic model of the Milky Way galaxy (see Sec.~\ref{sec:methods}) and for three values of the IMF index for massive stars ($>1\Msun$): $\Gamma=-2.3$ (standard), $-1.9$ (flat), and $-2.7$ (steep). $n_{\rm sl}$ is the estimated number of self-lensing sources detected during the duration of the survey. $f_{1,1}$ is the fraction of self-lensing binaries for which the expected number of flares, as well as the observations during one flare, are both higher than $1$. $f_{\rm NS}$ is the fraction of self-lensing binaries with NS lenses. $m_{\rm source,max}$ are the maximal values of the pre-magnified apparent magnitude of the source stars whose flares are observable and and $\mu_{\rm sl,max}$ is the highest magnification in the sample. The errors were calculated using the bootstrap method and 100 re-samplings of the population synthesis results.} 
    \label{tab:estimations}
\end{table*}

Our main results, showing the predicted number and nature of SL systems as observed by the chosen surveys (ZTF, LSST, and TESS), are presented in Table~\ref{tab:estimations}. The highest number of detections is predicted for the Rubin-LSST; this is a result of LSST being planned to last longer ($10$ years) than the ZTF or TESS surveys ($5$ and $2$ years, respectively), which not only increases the number of data points for short period SL binaries ($P_{\rm orb}<2$ years), but also the probability of detecting longer-period binaries ($P_{\rm orb}>>1\yr\Rightarrow P_{\rm obs}\propto t_{\rm survey}$). In addition, LSST has the highest apparent magnitude limit ($m_{\rm lim}=24/24.55$ for $5\sigma/3\sigma$) and nearly the entire Galactic stellar mass is localised inside its field-of-view ($\sim90\%$ of $M_{\rm MW}$). On the other end is TESS, which, despite its supreme cadence increasing the detection probability of very short events ($\tau_{\rm eff}\lesssim1$ day), and all-sky coverage, has the lowest predicted number of detections due to its much fainter apparent magnitude limit. 

For all instruments, the lenses producing the detected SL flares are typically (in $85$--$97\%$ of cases) NSs. Progenitors of NSs are more common in ZAMS populations than the progenitors of BHs due to the steepness of the high-mass end of the IMF. Additionally, low-mass stars, which are observable with these instruments due to their high apparent magnitude limits, are the typical companions to NSs. Binaries in which a BH progenitor is bound to a low mass star are not only rarer on ZAMS \citep{Wiktorowicz1911}, but also seldom survive the common envelope event, or their separations are large ($\gtrsim1000\Rsun$); as a result both the recurrence time is long and the probability that the impact parameters is $b\lesssim1$ (i.e. the event is included in our sample) is much smaller.

We have found that the majority ($\gtrsim99\%$) of the observed NSs and BHs in SL binaries are expected to be pristine, i.e. they have not been altered by any subsequent mass accretion since their formation. Finding and probing such objects (and especially the black holes as they will still have their natal spin) provides insights into the supernova mechanism \citep[e.g. the coupling of angular momentum to the ejecta -- e.g.][]{Fuller1908} and the connection between binary parameters and natal kicks. In this context, SL binaries can be more informative than X-ray binaries in which compact objects accrete mass and, therefore, change mass by up to several $\Msun$ \citep[e.g.][]{Wiktorowicz1911}. 

The maximum apparent magnitudes of source stars in our detected SL binaries ($m_{\rm source,max}$, Table~\ref{tab:estimations}), are typically fainter (up to $\sim8$--$9$ mag) than the apparent magnitude limits of the instruments (Table~\ref{tab:instruments}). Such a situation results from the fact that we seek to detect the flare and not necessarily the source star itself. The apparent magnitude of a flare ($m_{\rm flare}$) is related to the the apparent magnitude of the source star ($m_{\rm source}$) as
\begin{equation}
    m_{\rm flare}=m_{\rm source}+2.5\log_{10}(\mu_{\rm sl}-1).
\end{equation}
\noindent During the most extreme events ($\mu_{\rm sl}\approx1000$), the flare can reach the detection limit of an instrument even for very faint and unobservable source stars ($m_{\rm source}\gtrsim20$--$30$ mag).

The majority of the predicted SL sources have small magnifications ($\gtrsim89\%$ have $\mu_{\rm sl}-1<0.01$), which means that the source star is comparable to or larger than the Einstein radius. These are mainly massive stars with large luminosities which are observable (as are their relatively weak flares) out to larger distances, and mostly found in the vicinity of the Bulge, which significantly increases the fraction of the Galaxy where they can reside ($f_{\rm d}$), and consequently also $P_{\rm obs}$. The maximum magnification of our SL binaries is $\mu_{\rm sl}>100$ (see Table~\ref{tab:estimations}) in some cases, but the fraction of such highly magnified events in the sample is negligible ($\lll1\%$). 

\begin{figure*}
	\includegraphics[width=\textwidth]{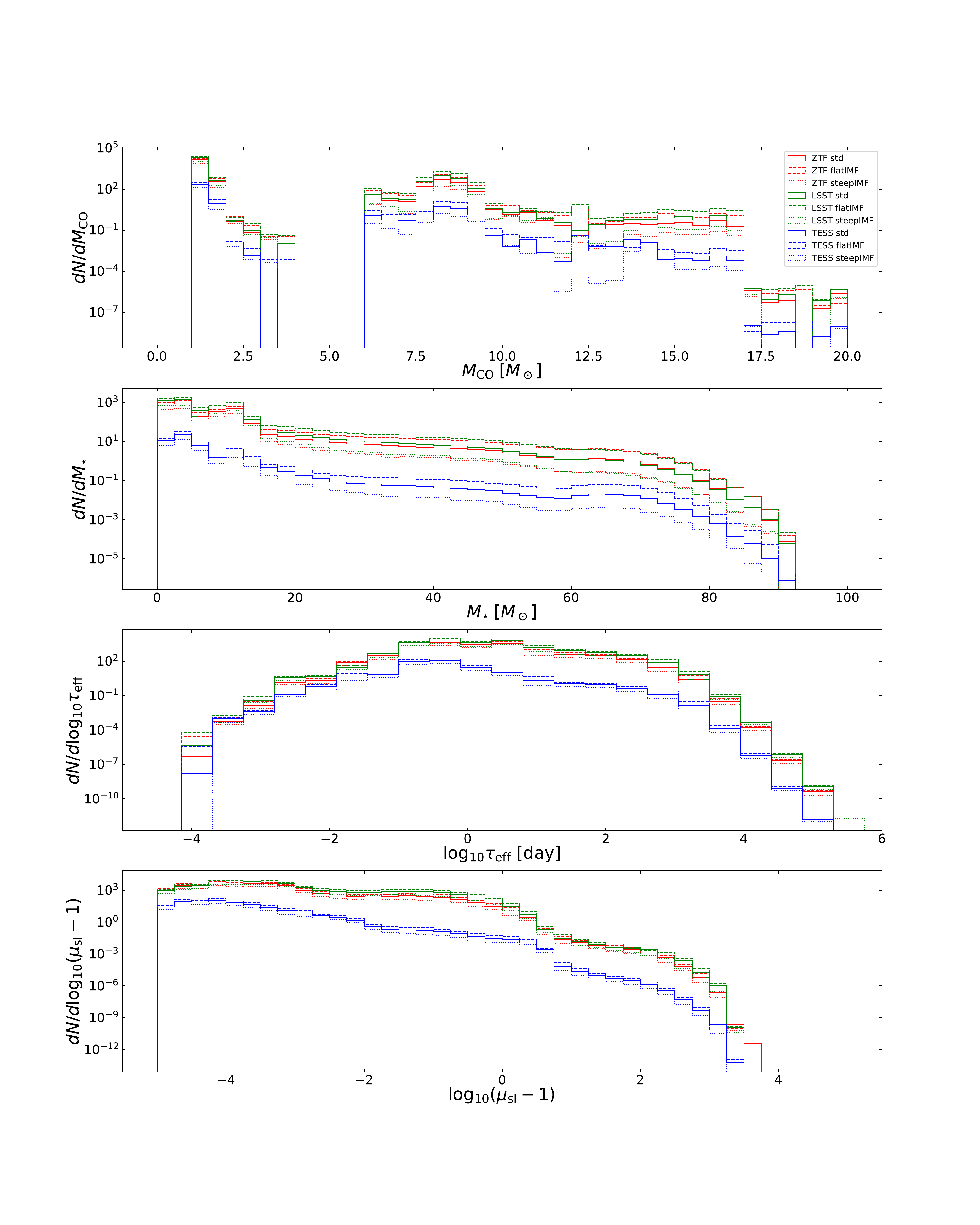}
    \caption{Distributions of compact object (lens) mass ($M_{\rm CO}$), companion (source star) mass ($M_{\star}$), the effective crossing time ($\tau_{\rm eff}$), and the relative luminosity of the SL flare ($\mu_{\rm sl}-1$) for three instruments (ZTF, LSST, and TESS) and three shapes of the IMF for massive ($M>1\Msun$) stars: std ($\Gamma=-2.3$), flatIMF ($\Gamma=-1.9$), and steepIMF ($\Gamma=-2.7$). See text for details.}
    \label{fig:distrs}
\end{figure*}

Figure~\ref{fig:distrs} presents the distributions for the parameters of SL binaries, and a comparison between different IMF slopes. The upper plot presents the lens masses; compact objects with masses below $2.5\Msun$ are assumed to be NSs, whereas those with masses above $2.5\Msun$ are BHs. The gap between $3$ and $5\Msun$ is an observational and theoretical feature \citep[e.g.][although see the recent LIGO results of \citeauthor{Abbott2006} \citeyear{Abbott2006} and the results from micro-lensing surveys, e.g. \citeauthor{Wyrzykowski2004} \citeyear{Wyrzykowski2004}]{Belczynski1209} BHs with masses above $17\Msun$ can exist in low metallicity parts of the Galaxy (the halo and the thick disk), but the low fraction of total Galactic stellar mass located in these Galactic components decreases the observation probability to such a level that these objects don't appear in our samples. 

The crossing times for the systems we predict to observe are typically small, although events longer than $100$ days appear in all surveys and in LSST even $>1000$ day events are present. This means that, except for the TESS sample, most of the events will have a very sparse coverage ($\sim1$ detection point); only  long events ($\tau_{\rm eff}\gtrsim10$ day) will have a more complete coverage. On the other hand, most of the self-lensing events within the TESS sample are expected to be covered more thoroughly; for example a $20$ minute event will already have $\sim10$ detection points. Such repeat events are extremely valuable as they allow for targeted follow-up observations (e.g. astrometry) and are immediately distinguishable from micro-lensing.

The lower panel of Figure~\ref{fig:distrs} shows that only a small fraction ($\lesssim0.2\%$) of SL events will have a flare as bright or brighter than the source star and actually the majority of events ($\gtrsim89\%$) are very weak with $\mu_{\rm sl}\lesssim1.01$. The latter can be observed only if the companion star is itself very luminous.

\begin{figure*}
	\includegraphics[width=\textwidth]{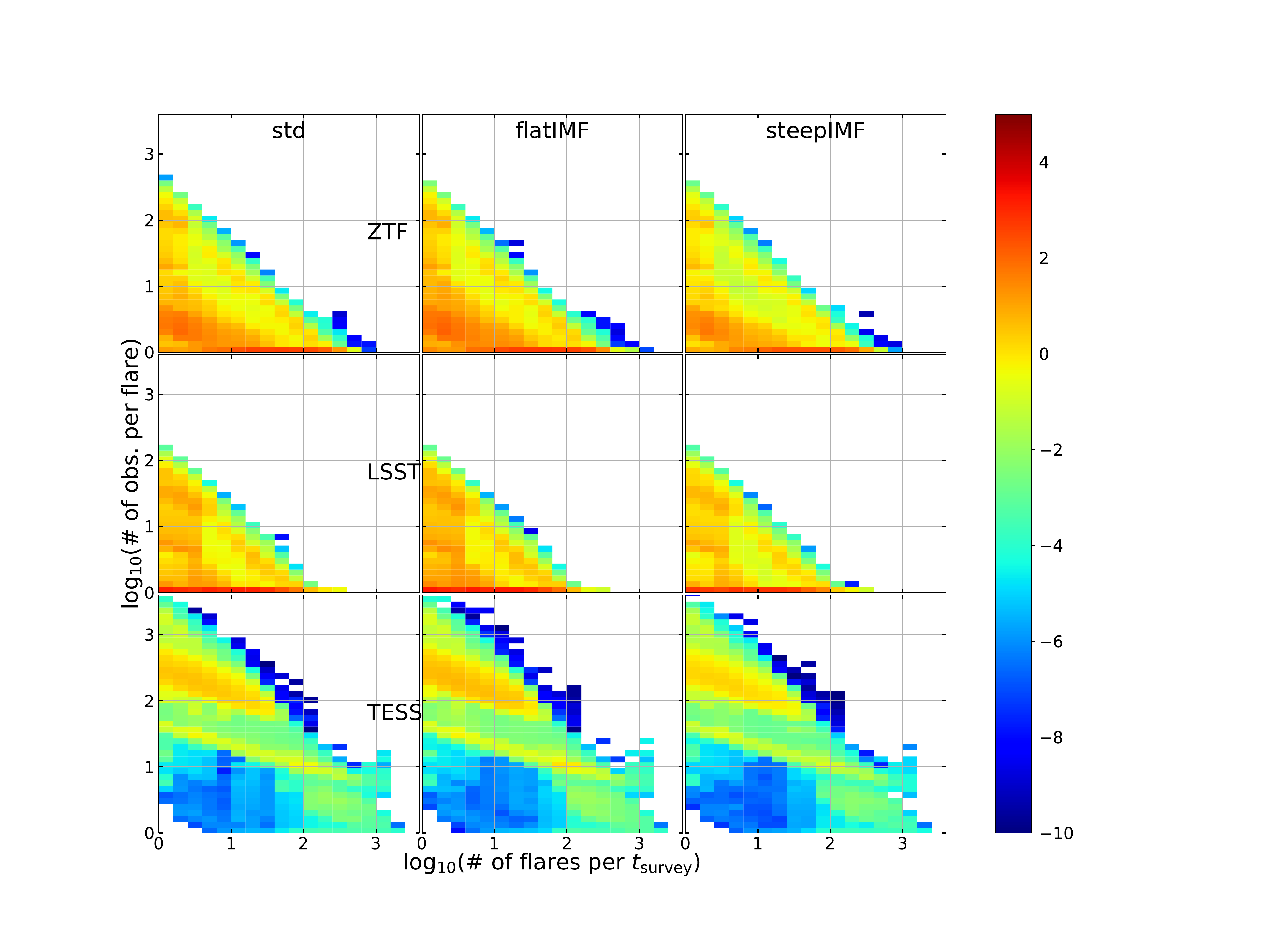}
    \caption{Distributions of number of flares ($N_{\rm rec}$) expected from a source (during each respective survey) versus number of data points per flare ($N_{\rm points}$) as a function of IMF slope.}
    \label{fig:rec_np}
\end{figure*}

\begin{table*}
    \label{tab:rec_np}
	\centering
    \begin{tabular}{ll|ccc|ccc|ccc}
        \hline
        \multicolumn{2}{l|}{Instrument \textbackslash\ Model}  & & std & & & flat & & & steep & \\
 & $N_{\rm rec}$ \textbackslash\ $N_{\rm points}$ & $>1$ & $>10$ & $>100$ & $>1$ & $>10$ & $>100$ & $>1$ & $>10$ & $>100$ \\
        \hline 
 & $>1$ & \sci{1.8}{3} & \sci{4.4}{1} & $7.7$ & \sci{2.5}{3} & \sci{6.0}{1} & $9.5$ & \sci{9.8}{2} & \sci{2.3}{1} & $4.0$ \\
ZTF & $>10$ & \sci{1.2}{3} & $2.1$ & -- & \sci{1.6}{3} & $2.6$ & -- & \sci{6.4}{2} & $1.3$ & -- \\
 & $>100$ & \sci{2.3}{2} & -- & -- & \sci{3.3}{2} & -- & -- & \sci{1.3}{2} & -- & -- \\
 & $>1$ & \sci{8.0}{3} & \sci{1.2}{2} & \sci{1.2}{-1} & \sci{1.1}{4} & \sci{1.6}{2} & \sci{2.1}{-1} & \sci{4.1}{3} & \sci{6.7}{1} & \sci{5.5}{-2} \\
LSST & $>10$ & \sci{3.2}{3} & \sci{3.5}{-2} & -- & \sci{4.2}{3} & \sci{4.7}{-2} & -- & \sci{1.7}{3} & \sci{1.8}{-2} & -- \\
 & $>100$ & $5.2$ & -- & -- & $4.0$ & -- & -- & $2.3$ & -- & -- \\
 & $>1$ & \sci{8.7}{1} & \sci{8.7}{1} & \sci{7.5}{1} & \sci{1.2}{2} & \sci{1.2}{2} & \sci{1.1}{2} & \sci{4.6}{1} & \sci{4.6}{1} & \sci{4.0}{1} \\
TESS & $>10$ & \sci{1.8}{1} & \sci{1.8}{1} & $9.3$ & \sci{2.8}{1} & \sci{2.7}{1} & \sci{1.5}{1} & $9.5$ & $9.3$ & $4.9$ \\
 & $>100$ & \sci{8.8}{-1} & \sci{3.1}{-1} & \sci{2.1}{-10} & $1.7$ & \sci{6.8}{-1} & \sci{7.3}{-10} & \sci{3.1}{-1} & \sci{8.5}{-2} & \sci{8.1}{-11} \\
        \hline
    \end{tabular}
    \caption{Expected number of detection in relations to recurrence, i.e. number of flares observed during the duration of a survey ($N_{\rm rec}$) and data coverage, i.e. the number of observations during a flare ($N_{\rm points}$).} 
\end{table*}

Our results indicate that, for $60$--$90\%$ of SL binaries, repeat events can be observed during each survey and up to $12\%$ have at least $100$ flares. However, the majority of the SL sources we predict we should observe ($\gtrsim1000$) will be hard to detect due to the typically low number of repeat flares during $t_{\rm survey}$ ($N_{\rm rec}$) or the low number of data points per flare ($N_{\rm points}$). See Figure~\ref{fig:rec_np} and Table~\ref{tab:rec_np}. The latter can be improved upon through stacking observations of several flares or spectroscopic observations of the companion's motion, if available, which can help to model the entire light curve even if some flares are missed. We note the persistent issue that single flare detections (especially, when $R_\star<<R_{\rm E}$) can be easily misinterpreted as micro-lensing events. Additionally, when the data coverage of a flare is low, misinterpretation as other kinds of outburst (e.g. stellar flares) is also possible. 

Figure~\ref{fig:rec_np} illustrates histograms of the expected number of systems with specified limits for the number of SL flare recurrences ($N_{\rm rec}=\min(\tau_{\rm eff}/t_{\rm c},1) t_{\rm survey}/P_{\rm orb}$, except for those systems calculated using `brute force', see Section~\ref{sec:methods}), and expected data coverage of the flares ($N_{\rm points}=\tau_{\rm eff}/t_{\rm c}$) for all of the instruments we are considering and the three IMF models (the corresponding numbers are provided in Table~\ref{tab:rec_np}). Non-integer values of $N_{\rm rec}$ or $N_{\rm points}$ should be interpreted as the expected number of systems, e.g. $N_{\rm rec}\leq1$ means a probability of observing one flare during $t_{\rm survey}$ is $P_{\rm obs}=N_{\rm rec}$. The histograms show a inverse proportionality between the estimated number of detections and $N_{\rm rec}$ or $N_{\rm points}$. Actually, $N_{\rm rec}$ and $N_{\rm points}$ are not independent, because $N_{\rm rec}\sim P_{\rm orb}^{-1}\sim v_{\rm orb}\sim \tau_{\rm eff}^{-1}\sim N_{\rm points}^{-1}$. 

The lack of data in the top right-hand corners of each panel in Figure~\ref{fig:rec_np} is related to $\tau_{\rm eff}\approx \frac{1}{2} P_{\rm orb}$, i.e. when the companion star is nearly filling its Roche lobe, which can be as large as $\sim0.8a$ for an extreme case ($M_{\rm comp}>>M_{\rm CO}$), so  it is the largest possible size of a Roche lobe for a non-interacting binary; for any particular $P_{\rm orb}$, this yields the maximum number of observations of a flare. In such situations, the larger the $P_{\rm orb}$, the larger the number of possible observations of the flare ($N_{\rm points}\sim 0.5P_{\rm orb}$) which explains the observed anti-correlation. Finally, the cut-off at $N_{\rm points}=1$ is imposed by the authors to show only systems in which all consecutive flares are visible. If $N_{\rm points}$ is lower than one, only a fraction of these flares will have an observational point, which may lead to misinterpretation of $N_{\rm rec}$.

\begin{figure*}
	\includegraphics[width=\textwidth]{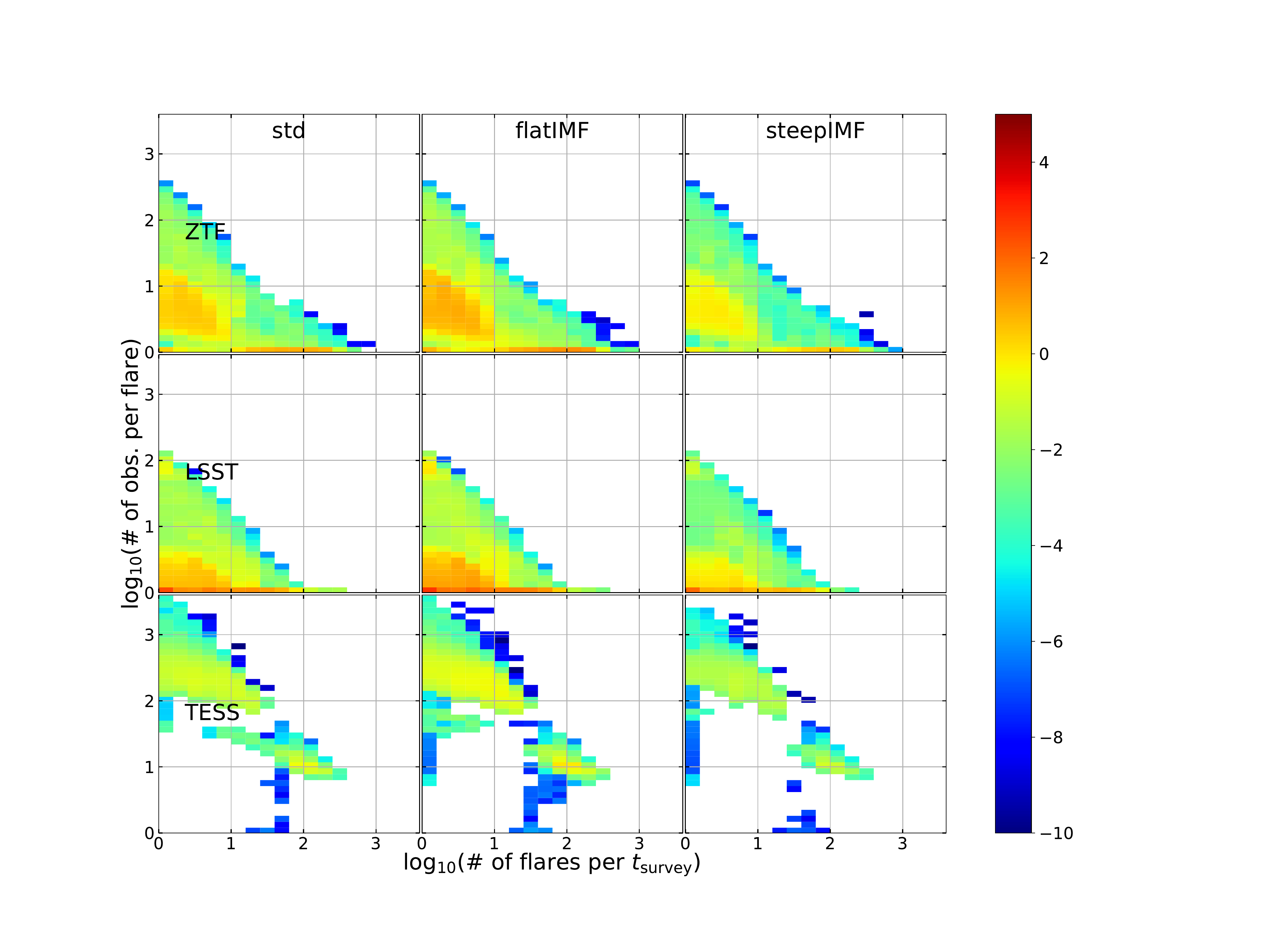}
    \caption{Same as Figure~\ref{fig:rec_np}, but for BH lenses only.}
    \label{fig:rec_np_bh}
\end{figure*}

\begin{figure*}
	\includegraphics[width=\textwidth]{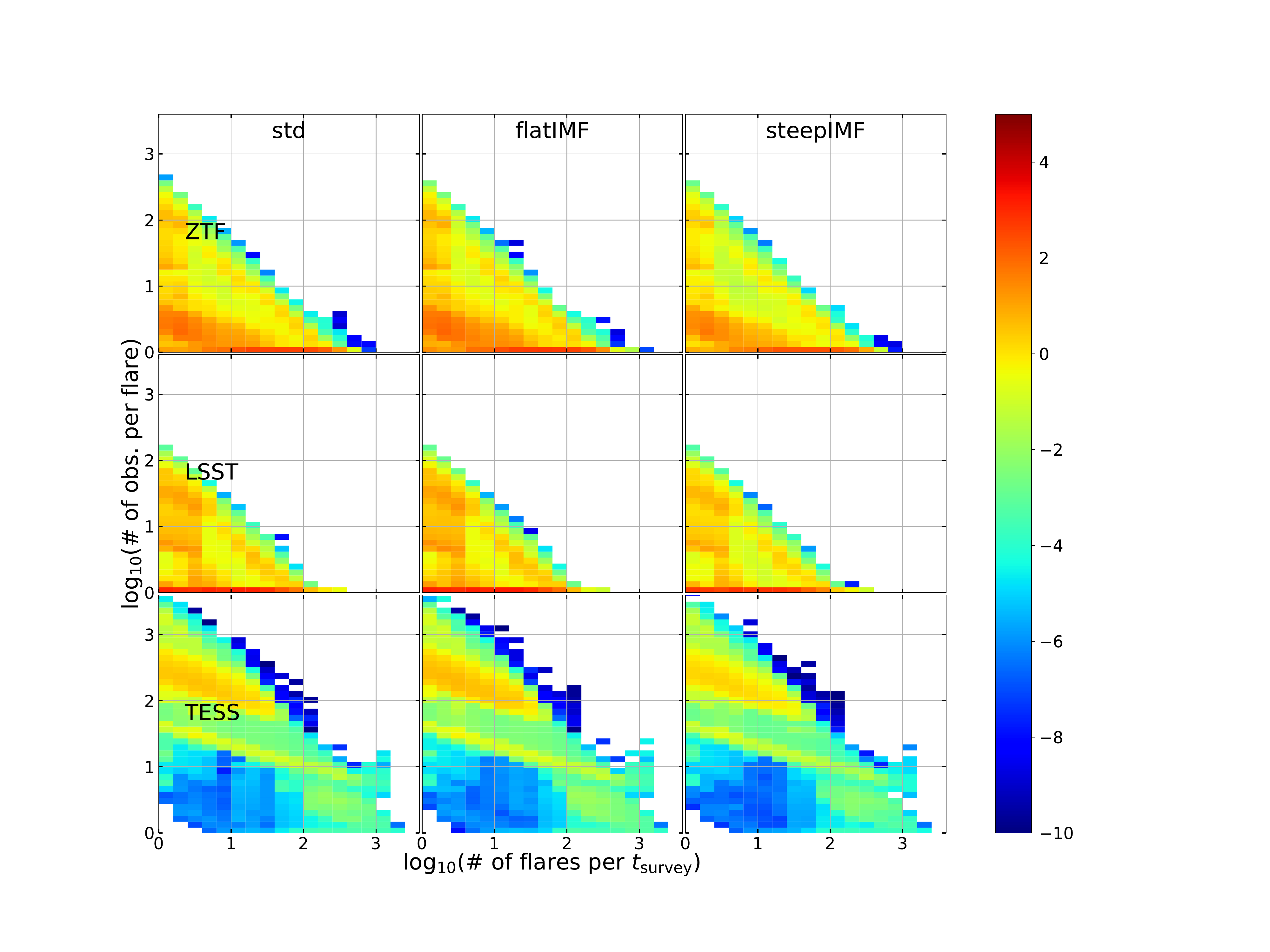}
    \caption{Same as Figure~\ref{fig:rec_np}, but for NS lenses only.}
    \label{fig:rec_np_ns}
\end{figure*}

In Figures~\ref{fig:rec_np_bh} and \ref{fig:rec_np_ns}, we present a similar collection of plots as in Figure~\ref{fig:rec_np}, but only for SL binaries with BH or NS lenses, respectively. Apart from the dominance of NS lenses in the population of SL binaries ($>85\%$; see Table~\ref{tab:estimations}), there are no significant differences between BH lenses and NS lenses as far as $N_{\rm rec}$ and $N_{\rm points}$ are concerned.

\section{Discussion \& Conclusions}

In this study we present a complete theoretical model of SL emission for binaries harbouring a BH or a NS. We have shown that observations of repeating SL flares are a promising way of detecting non-interacting compact object binaries. Using the comprehensive model of the Milky Way galaxy developed in \citet{Wiktorowicz2006} and the database of BHs in stellar populations \citep{Wiktorowicz1911} extended for systems in which the compact object is a NS, we have calculated realistic predictions for three ongoing or planned large-area optical surveys: ZTF GPS, TESS, and the Vera Rubin-LSST. Our results indicate that such current or future optical surveys may have the ability to detect hundreds to thousands of SL binary systems.

Our main results are summarised in Table~\ref{tab:estimations}. We show that the highest number of SL events ($\sim5,300$ -- $14,000$) are predicted to be observed by LSST, as a result of the much longer duration of this survey compared to ZTF or TESS, and a very high apparent magnitude detection limit. Conversely, TESS -- which will observe every part of the sky for only 27 days, with overlapping parts of up to one year -- is predicted to detect only $\sim60$--$170$ SL sources, a significant fraction of which will be post-interaction and in which mass transfer has already taken place. However, TESS, due to its supremely high cadence, is the preferred tool for providing the highest data coverage and many recurrences (e.g. 18 sources with $N_{\rm rec}>10$ and $N_{\rm point}>10$ in contrast to two and none such sources in ZTF and LSST samples, respectively)

We have found that the majority of the self-lensing flares will be very weak ($\mu_{\rm sl}-1<0.01$); this results from the fact that compact objects are typically accompanied by massive stars and massive stars are mostly located with the vicinity of the bulge. Consequently, the radius of the companion star is typically much larger than the Einstein radius, which results in small magnifications (see Fig.~\ref{fig:magnification} and related text). For such a relatively small flare to be detectable, the intrinsic luminosity of the companion star needs to be large so even a small magnification will be visible above the noise, or many flares need to be stacked together. As massive stars are rare due to their short lifetimes, only surveys covering a large fraction of the galaxy can obtain a sizable sample. ZTF and LSST are able to surmount this problem due to their fainter magnitude limit ($m_{\rm lim}=21$ and $24$, respectively) as compared to TESS ($m_{\rm lim}=12$). We note the flares with $\mu_{\rm sl}\approx1$ may be hard to distinguish from other sources of flux variability even if the data coverage is moderate. As a consequence, many flares may be lost during the observations. These observation-related issues are not a part of this study, so our results in this context should be perceived as an estimated upper limit on the number of detectable SL sources. Nonetheless, we predict a significant number (e.g. $18$ for TESS) of systems which have $N_{\rm rec}>10$ and $N_{\rm points}>10$, so should be relatively easy to detect. What is more, reducing the constrains to $N_{\rm rec}>1$ and $N_{\rm points}>10$ we obtain a prediction of $120$ detections with LSST ($44$ with ZFT and $87$ with TESS), which is already comparable in size to the known sample of XRBs in our Galaxy. However, only $\sim1\%$ of those binaries harbour BHs.

Some SL binaries with long orbital periods ($P_{\rm orb}\gtrsim t_{\rm survey}$) may be detected in serendipitous fashion when the flare occurs precisely at the moment of observation. However, a single SL flare is hard to distinguish from a micro-lensing event. Therefore, the only way to discriminate these two types of events in survey data is to look for recurrence, which is possible only if $P_{\rm orb}<t_{\rm survey}$. To explore this further, we have calculated the expected recurrence and, simultaneously, the data coverage of flares, which can help distinguish them from other types of stellar flares through lightcurve fitting. We provide the results in Figure~\ref{fig:rec_np} and Table~\ref{tab:rec_np}. From our results it is evident that the number of detected SL binaries drops very quickly with required number of recurrences, especially if a data coverage higher than $10$ points per flare is required. This can lead to the possibility of confusion with other events, and the requirement for advanced filtering and modelling.

\citet{Masuda1910} have already performed a proof-of-concept study of detecting non-interacting BH binaries using SL. With a simplified population synthesis model they predict $\sim10$ BHs to be detected by TESS through this method. Our study significantly expands on their work by utilising a comprehensive population synthesis code \citep[\startrack][]{Belczynski0801}, a realistic model for the Galaxy, consideration of other major surveys (ZTF and LSST) and the inclusion of NS systems. Our results show that up to $\sim300$ BHs could be found in TESS data (and considerably more in the other surveys), which is roughly consistent with their results. 

As we show in our work, the majority of NS and BH lenses in SL binaries ($\gtrsim99\%$) are pristine compact objects, i.e. they have not accreted mass since their formation, therefore, the distribution of the properties of these compact objects can provide us with a much better -- perhaps unique -- understanding of the supernova process than the one obtained from X-ray binaries (XRBs). In addition, SL binaries represent a much broader sub-population of binaries than XRBs which always evolve through a common envelope phase \citep[see e.g.][]{Wiktorowicz1409} and whose life-times are much shorter. Especially, observations of XRBs may be biased against systems harbouring the most massive BHs for which orbital periods and, therefore, recurrence times are much longer \citep{Jonker2104}. In the case of SL binaries this problem will be mitigated by the larger size of the expected observational sample.

Observations of SL binaries provide an important new means to investigate the evolution of non-interacting binaries with one compact component. Detecting and modelling the repeating SL flares will shed light on evolutionary phases, which although generating little continuous emission, are typically the longest evolutionary phases of BH or NS binaries. In this paper we showed that ongoing and future optical surveys can provide statistically reliable samples of SL sources harbouring BHs and NSs that will not be readily accessible via other techniques \citep[e.g. low angular momentum accretion --][]{Proga0308}. In future we plan to significantly extend this study by testing a wider variety of evolutionary models and providing predictions for other instruments. We will also include predictions based on ray-tracing models for self-lensing in various metrics as well as including processes which make the surface luminosity of the source star non-homogeneous (such as limb and gravitational darkening).

\bigskip
\section*{Data availability}
New data generated during the course of this project is publicly available at \url{https://universeathome.pl/universe/self_lensing.php}.
\bigskip
\section*{Acknowledgements}

We are thankful to thousands of volunteers, who took part in the {\it Universe@Home} project\footnote{https://universeathome.pl/} and significantly sped up the simulations. GW is partly supported by the President’s International Fellowship Initiative (PIFI) of the Chinese Academy of Sciences under grant no.2018PM0017 and by the Strategic Priority Research Program of the Chinese Academy of Science Multi-waveband Gravitational Wave Universe (Grant No. XDB23040000). This work is partly supported by the National Natural Science Foundation of China (Grant No. 11690024, 11873056, and 11991052) and the National Key Program for Science and Technology Research and Development (Grant No. 2016YFA0400704). AI acknowledges support from the Royal Society.

\bibliographystyle{mnras}
\bibliography{main}

\begin{thebibliography}{}
\makeatletter
\relax
\def\mn@urlcharsother{\let\do\@makeother \do\$\do\&\do\#\do\^\do\_\do\%\do\~}
\def\mn@doi{\begingroup\mn@urlcharsother \@ifnextchar [ {\mn@doi@}
  {\mn@doi@[]}}
\def\mn@doi@[#1]#2{\def\@tempa{#1}\ifx\@tempa\@empty \href
  {http://dx.doi.org/#2} {doi:#2}\else \href {http://dx.doi.org/#2} {#1}\fi
  \endgroup}
\def\mn@eprint#1#2{\mn@eprint@#1:#2::\@nil}
\def\mn@eprint@arXiv#1{\href {http://arxiv.org/abs/#1} {{\tt arXiv:#1}}}
\def\mn@eprint@dblp#1{\href {http://dblp.uni-trier.de/rec/bibtex/#1.xml}
  {dblp:#1}}
\def\mn@eprint@#1:#2:#3:#4\@nil{\def\@tempa {#1}\def\@tempb {#2}\def\@tempc
  {#3}\ifx \@tempc \@empty \let \@tempc \@tempb \let \@tempb \@tempa \fi \ifx
  \@tempb \@empty \def\@tempb {arXiv}\fi \@ifundefined
  {mn@eprint@\@tempb}{\@tempb:\@tempc}{\expandafter \expandafter \csname
  mn@eprint@\@tempb\endcsname \expandafter{\@tempc}}}

\bibitem[\protect\citeauthoryear{{Abbott} et~al.,}{{Abbott}
  et~al.}{2020}]{Abbott2006}
{Abbott} R.,  et~al., 2020, \mn@doi [\apjl] {10.3847/2041-8213/ab960f}, \href
  {https://ui.adsabs.harvard.edu/abs/2020ApJ...896L..44A} {896, L44}

\bibitem[\protect\citeauthoryear{{Agol}}{{Agol}}{2003}]{Agol0309}
{Agol} E.,  2003, \mn@doi [\apj] {10.1086/376833}, \href
  {https://ui.adsabs.harvard.edu/abs/2003ApJ...594..449A} {594, 449}

\bibitem[\protect\citeauthoryear{{Belczynski}, {Kalogera}, {Rasio}, {Taam},
  {Zezas}, {Bulik}, {Maccarone}  \& {Ivanova}}{{Belczynski}
  et~al.}{2008}]{Belczynski0801}
{Belczynski} K.,  {Kalogera} V.,  {Rasio} F.~A.,  {Taam} R.~E.,  {Zezas} A.,
  {Bulik} T.,  {Maccarone} T.~J.,   {Ivanova} N.,  2008, \mn@doi [\apjs]
  {10.1086/521026}, \href
  {https://ui.adsabs.harvard.edu/abs/2008ApJS..174..223B} {174, 223}

\bibitem[\protect\citeauthoryear{{Belczynski}, {Wiktorowicz}, {Fryer}, {Holz}
  \& {Kalogera}}{{Belczynski} et~al.}{2012}]{Belczynski1209}
{Belczynski} K.,  {Wiktorowicz} G.,  {Fryer} C.~L.,  {Holz} D.~E.,   {Kalogera}
  V.,  2012, \mn@doi [\apj] {10.1088/0004-637X/757/1/91}, \href
  {https://ui.adsabs.harvard.edu/abs/2012ApJ...757...91B} {757, 91}

\bibitem[\protect\citeauthoryear{{Belczynski} et~al.,}{{Belczynski}
  et~al.}{2020}]{Belczynski2004}
{Belczynski} K.,  et~al., 2020, \mn@doi [\aap] {10.1051/0004-6361/201936528},
  \href {https://ui.adsabs.harvard.edu/abs/2020A&A...636A.104B} {636, A104}

\bibitem[\protect\citeauthoryear{{Casares}, {Negueruela}, {Rib{\'o}}, {Ribas},
  {Paredes}, {Herrero}  \& {Sim{\'o}n-D{\'\i}az}}{{Casares}
  et~al.}{2014}]{Casares1401}
{Casares} J.,  {Negueruela} I.,  {Rib{\'o}} M.,  {Ribas} I.,  {Paredes} J.~M.,
  {Herrero} A.,   {Sim{\'o}n-D{\'\i}az} S.,  2014, \mn@doi [\nat]
  {10.1038/nature12916}, \href
  {https://ui.adsabs.harvard.edu/abs/2014Natur.505..378C} {505, 378}

\bibitem[\protect\citeauthoryear{{Chen} et~al.,}{{Chen}
  et~al.}{2019}]{Chen1912}
{Chen} Y.,  et~al., 2019, \mn@doi [\aap] {10.1051/0004-6361/201936612}, \href
  {https://ui.adsabs.harvard.edu/abs/2019A&A...632A.105C} {632, A105}

\bibitem[\protect\citeauthoryear{{D'Orazio} \& {Di Stefano}}{{D'Orazio} \& {Di
  Stefano}}{2018}]{DOrazio1803}
{D'Orazio} D.~J.,  {Di Stefano} R.,  2018, \mn@doi [\mnras]
  {10.1093/mnras/stx2936}, \href
  {https://ui.adsabs.harvard.edu/abs/2018MNRAS.474.2975D} {474, 2975}

\bibitem[\protect\citeauthoryear{{D'Orazio} \& {Di Stefano}}{{D'Orazio} \& {Di
  Stefano}}{2020}]{DOrazio2001}
{D'Orazio} D.~J.,  {Di Stefano} R.,  2020, \mn@doi [\mnras]
  {10.1093/mnras/stz3086}, \href
  {https://ui.adsabs.harvard.edu/abs/2020MNRAS.491.1506D} {491, 1506}

\bibitem[\protect\citeauthoryear{{Ebrahimnejad Rahbari}, {Nouri-Zonoz}  \&
  {Rahvar}}{{Ebrahimnejad Rahbari} et~al.}{2005}]{EbrahimnejadRahbari0508}
{Ebrahimnejad Rahbari} H.,  {Nouri-Zonoz} M.,   {Rahvar} S.,  2005, arXiv
  e-prints, \href {https://ui.adsabs.harvard.edu/abs/2005astro.ph..8477E} {pp
  astro--ph/0508477}

\bibitem[\protect\citeauthoryear{{Fuller} \& {Ma}}{{Fuller} \&
  {Ma}}{2019}]{Fuller1908}
{Fuller} J.,  {Ma} L.,  2019, \mn@doi [\apjl] {10.3847/2041-8213/ab339b}, \href
  {https://ui.adsabs.harvard.edu/abs/2019ApJ...881L...1F} {881, L1}

\bibitem[\protect\citeauthoryear{{Gandhi}, {Rao}, {Charles}, {Belczynski},
  {Maccarone}, {Arur}  \& {Corral-Santana}}{{Gandhi} et~al.}{2020}]{Gandhi2005}
{Gandhi} P.,  {Rao} A.,  {Charles} P.~A.,  {Belczynski} K.,  {Maccarone} T.~J.,
   {Arur} K.,   {Corral-Santana} J.~M.,  2020, \mn@doi [\mnras]
  {10.1093/mnrasl/slaa081}, \href
  {https://ui.adsabs.harvard.edu/abs/2020MNRAS.496L..22G} {496, L22}

\bibitem[\protect\citeauthoryear{{Gould}}{{Gould}}{1995}]{Gould9506}
{Gould} A.,  1995, \mn@doi [\apj] {10.1086/175812}, \href
  {https://ui.adsabs.harvard.edu/abs/1995ApJ...446..541G} {446, 541}

\bibitem[\protect\citeauthoryear{{Gould} \& {Salim}}{{Gould} \&
  {Salim}}{2002}]{Gould0606}
{Gould} A.,  {Salim} S.,  2002, \mn@doi [\apj] {10.1086/340435}, \href
  {https://ui.adsabs.harvard.edu/abs/2002ApJ...572..944G} {572, 944}

\bibitem[\protect\citeauthoryear{{Harper}, {Brown}, {Guinan}, {O'Gorman},
  {Richards}, {Kervella}  \& {Decin}}{{Harper} et~al.}{2017}]{Harper1707}
{Harper} G.~M.,  {Brown} A.,  {Guinan} E.~F.,  {O'Gorman} E.,  {Richards}
  A.~M.~S.,  {Kervella} P.,   {Decin} L.,  2017, \mn@doi [\aj]
  {10.3847/1538-3881/aa6ff9}, \href
  {https://ui.adsabs.harvard.edu/abs/2017AJ....154...11H} {154, 11}

\bibitem[\protect\citeauthoryear{{Ingram}, {Motta}, {Aigrain}  \&
  {Karastergiou}}{{Ingram} et~al.}{2021}]{Ingram2021}
{Ingram} A.,  {Motta} S.~E.,  {Aigrain} S.,   {Karastergiou} A.,  2021, \mn@doi
  [\mnras] {10.1093/mnras/stab609}, \href
  {https://ui.adsabs.harvard.edu/abs/2021MNRAS.503.1703I} {503, 1703}

\bibitem[\protect\citeauthoryear{{Ivezi{\'c}} et~al.,}{{Ivezi{\'c}}
  et~al.}{2019}]{Ivezic1903}
{Ivezi{\'c}} {\v{Z}}.,  et~al., 2019, \mn@doi [\apj]
  {10.3847/1538-4357/ab042c}, \href
  {https://ui.adsabs.harvard.edu/abs/2019ApJ...873..111I} {873, 111}

\bibitem[\protect\citeauthoryear{{Johnson}, {Gandhi}, {Chapman}, {Moreau},
  {Charles}, {Clarkson}  \& {Hill}}{{Johnson} et~al.}{2019}]{Johnson1903}
{Johnson} M. A.~C.,  {Gandhi} P.,  {Chapman} A.~P.,  {Moreau} L.,  {Charles}
  P.~A.,  {Clarkson} W.~I.,   {Hill} A.~B.,  2019, \mn@doi [\mnras]
  {10.1093/mnras/sty3466}, \href
  {https://ui.adsabs.harvard.edu/abs/2019MNRAS.484...19J} {484, 19}

\bibitem[\protect\citeauthoryear{{Jonker}, {Kaur}, {Stone}  \&
  {Torres}}{{Jonker} et~al.}{2021}]{Jonker2104}
{Jonker} P.~G.,  {Kaur} K.,  {Stone} N.,   {Torres} M. A.~P.,  2021, arXiv
  e-prints, \href {https://ui.adsabs.harvard.edu/abs/2021arXiv210403596J} {p.
  arXiv:2104.03596}

\bibitem[\protect\citeauthoryear{{Kawahara}, {Masuda}, {MacLeod}, {Latham},
  {Bieryla}  \& {Benomar}}{{Kawahara} et~al.}{2018}]{Kawahara1803}
{Kawahara} H.,  {Masuda} K.,  {MacLeod} M.,  {Latham} D.~W.,  {Bieryla} A.,
  {Benomar} O.,  2018, \mn@doi [\aj] {10.3847/1538-3881/aaaaaf}, \href
  {https://ui.adsabs.harvard.edu/abs/2018AJ....155..144K} {155, 144}

\bibitem[\protect\citeauthoryear{{Kruse} \& {Agol}}{{Kruse} \&
  {Agol}}{2014}]{Kruse1404}
{Kruse} E.,  {Agol} E.,  2014, \mn@doi [Science] {10.1126/science.1251999},
  \href {https://ui.adsabs.harvard.edu/abs/2014Sci...344..275K} {344, 275}

\bibitem[\protect\citeauthoryear{{Masci} et~al.,}{{Masci}
  et~al.}{2019}]{Masci1901}
{Masci} F.~J.,  et~al., 2019, \mn@doi [\pasp] {10.1088/1538-3873/aae8ac}, \href
  {https://ui.adsabs.harvard.edu/abs/2019PASP..131a8003M} {131, 018003}

\bibitem[\protect\citeauthoryear{{Masuda} \& {Hotokezaka}}{{Masuda} \&
  {Hotokezaka}}{2019}]{Masuda1910}
{Masuda} K.,  {Hotokezaka} K.,  2019, \mn@doi [\apj]
  {10.3847/1538-4357/ab3a4f}, \href
  {https://ui.adsabs.harvard.edu/abs/2019ApJ...883..169M} {883, 169}

\bibitem[\protect\citeauthoryear{{Masuda}, {Kawahara}, {Latham}, {Bieryla},
  {Kunitomo}, {MacLeod}  \& {Aoki}}{{Masuda} et~al.}{2019}]{Masuda1908}
{Masuda} K.,  {Kawahara} H.,  {Latham} D.~W.,  {Bieryla} A.,  {Kunitomo} M.,
  {MacLeod} M.,   {Aoki} W.,  2019, \mn@doi [\apjl] {10.3847/2041-8213/ab321b},
  \href {https://ui.adsabs.harvard.edu/abs/2019ApJ...881L...3M} {881, L3}

\bibitem[\protect\citeauthoryear{{Proga} \& {Begelman}}{{Proga} \&
  {Begelman}}{2003}]{Proga0308}
{Proga} D.,  {Begelman} M.~C.,  2003, \mn@doi [\apj] {10.1086/375773}, \href
  {https://ui.adsabs.harvard.edu/abs/2003ApJ...592..767P} {592, 767}

\bibitem[\protect\citeauthoryear{{Remillard} \& {McClintock}}{{Remillard} \&
  {McClintock}}{2006}]{Remillard0609}
{Remillard} R.~A.,  {McClintock} J.~E.,  2006, \mn@doi [\araa]
  {10.1146/annurev.astro.44.051905.092532}, \href
  {https://ui.adsabs.harvard.edu/abs/2006ARA&A..44...49R} {44, 49}

\bibitem[\protect\citeauthoryear{{Ricker} et~al.,}{{Ricker}
  et~al.}{2014}]{Ricker1408}
{Ricker} G.~R.,  et~al., 2014, in {Oschmann} Jacobus~M. J.,  {Clampin} M.,
  {Fazio} G.~G.,   {MacEwen} H.~A.,  eds,  Society of Photo-Optical
  Instrumentation Engineers (SPIE) Conference Series Vol. 9143, Space
  Telescopes and Instrumentation 2014: Optical, Infrared, and Millimeter Wave.
  p. 914320 (\mn@eprint {arXiv} {1406.0151}), \mn@doi{10.1117/12.2063489}

\bibitem[\protect\citeauthoryear{{Ricker} et~al.,}{{Ricker}
  et~al.}{2015}]{Ricker1501}
{Ricker} G.~R.,  et~al., 2015, \mn@doi [Journal of Astronomical Telescopes,
  Instruments, and Systems] {10.1117/1.JATIS.1.1.014003}, \href
  {https://ui.adsabs.harvard.edu/abs/2015JATIS...1a4003R} {1, 014003}

\bibitem[\protect\citeauthoryear{{Robin}, {Reyl{\'e}}, {Derri{\`e}re}  \&
  {Picaud}}{{Robin} et~al.}{2003}]{Robin0310}
{Robin} A.~C.,  {Reyl{\'e}} C.,  {Derri{\`e}re} S.,   {Picaud} S.,  2003,
  \mn@doi [\aap] {10.1051/0004-6361:20031117}, \href
  {https://ui.adsabs.harvard.edu/abs/2003A&A...409..523R} {409, 523}

\bibitem[\protect\citeauthoryear{{Spitzer}}{{Spitzer}}{1978}]{Spitzer78}
{Spitzer} L.,  1978, {Physical processes in the interstellar medium},
  \mn@doi{10.1002/9783527617722.
}

\bibitem[\protect\citeauthoryear{{Sullivan} et~al.,}{{Sullivan}
  et~al.}{2015}]{Sullivan1508}
{Sullivan} P.~W.,  et~al., 2015, \mn@doi [\apj] {10.1088/0004-637X/809/1/77},
  \href {https://ui.adsabs.harvard.edu/abs/2015ApJ...809...77S} {809, 77}

\bibitem[\protect\citeauthoryear{{Turner}, {Kovtyukh}, {Usenko}  \&
  {Gorlova}}{{Turner} et~al.}{2013}]{Turner1301}
{Turner} D.~G.,  {Kovtyukh} V.~V.,  {Usenko} I.~A.,   {Gorlova} N.~I.,  2013,
  \mn@doi [\apjl] {10.1088/2041-8205/762/1/L8}, \href
  {https://ui.adsabs.harvard.edu/abs/2013ApJ...762L...8T} {762, L8}

\bibitem[\protect\citeauthoryear{{Wiktorowicz}, {Belczynski}  \&
  {Maccarone}}{{Wiktorowicz} et~al.}{2014}]{Wiktorowicz1409}
{Wiktorowicz} G.,  {Belczynski} K.,   {Maccarone} T.,  2014, in {de Grijs} R.,
  ed., Binary Systems, their Evolution and Environments. p.~37 (\mn@eprint
  {arXiv} {1312.5924})

\bibitem[\protect\citeauthoryear{{Wiktorowicz}, {Wyrzykowski}, {Chruslinska},
  {Klencki}, {Rybicki}  \& {Belczynski}}{{Wiktorowicz}
  et~al.}{2019}]{Wiktorowicz1911}
{Wiktorowicz} G.,  {Wyrzykowski} {\L}.,  {Chruslinska} M.,  {Klencki} J.,
  {Rybicki} K.~A.,   {Belczynski} K.,  2019, \mn@doi [\apj]
  {10.3847/1538-4357/ab45e6}, \href
  {https://ui.adsabs.harvard.edu/abs/2019ApJ...885....1W} {885, 1}

\bibitem[\protect\citeauthoryear{{Wiktorowicz}, {Lu}, {Wyrzykowski}, {Zhang},
  {Liu}, {Justham}  \& {Belczynski}}{{Wiktorowicz}
  et~al.}{2020}]{Wiktorowicz2006}
{Wiktorowicz} G.,  {Lu} Y.,  {Wyrzykowski} {\L}.,  {Zhang} H.,  {Liu} J.,
  {Justham} S.,   {Belczynski} K.,  2020, arXiv e-prints, \href
  {https://ui.adsabs.harvard.edu/abs/2020arXiv200608317W} {p. arXiv:2006.08317}

\bibitem[\protect\citeauthoryear{{Witt} \& {Mao}}{{Witt} \&
  {Mao}}{1994}]{Witt9408}
{Witt} H.~J.,  {Mao} S.,  1994, \mn@doi [\apj] {10.1086/174426}, \href
  {https://ui.adsabs.harvard.edu/abs/1994ApJ...430..505W} {430, 505}

\bibitem[\protect\citeauthoryear{{Wyrzykowski} \& {Mandel}}{{Wyrzykowski} \&
  {Mandel}}{2020}]{Wyrzykowski2004}
{Wyrzykowski} {\L}.,  {Mandel} I.,  2020, \mn@doi [\aap]
  {10.1051/0004-6361/201935842}, \href
  {https://ui.adsabs.harvard.edu/abs/2020A&A...636A..20W} {636, A20}

\bibitem[\protect\citeauthoryear{{Wyrzykowski} et~al.,}{{Wyrzykowski}
  et~al.}{2011}]{Wyrzykowski1110}
{Wyrzykowski} L.,  et~al., 2011, \mn@doi [\mnras]
  {10.1111/j.1365-2966.2011.19243.x}, \href
  {https://ui.adsabs.harvard.edu/abs/2011MNRAS.416.2949W} {416, 2949}

\bibitem[\protect\citeauthoryear{{van Haaften}, {Nelemans}, {Voss}, {van der
  Sluys}  \& {Toonen}}{{van Haaften} et~al.}{2015}]{vanHaaften1507}
{van Haaften} L.~M.,  {Nelemans} G.,  {Voss} R.,  {van der Sluys} M.~V.,
  {Toonen} S.,  2015, \mn@doi [\aap] {10.1051/0004-6361/201425303}, \href
  {https://ui.adsabs.harvard.edu/abs/2015A&A...579A..33V} {579, A33}

\bibitem[\protect\citeauthoryear{{van Leeuwen}}{{van
  Leeuwen}}{2007}]{vanLeeuwen0711}
{van Leeuwen} F.,  2007, \mn@doi [\aap] {10.1051/0004-6361:20078357}, \href
  {https://ui.adsabs.harvard.edu/abs/2007A&A...474..653V} {474, 653}

\makeatother
\end{thebibliography}

\end{document}